# Quantum sensing and imaging of spin-orbit-torque-driven spin dynamics in noncollinear antiferromagnet Mn$_3$Sn

Gerald Q. Yan[1,†], Senlei Li[1,†], Hanyi Lu[1], Mengqi Huang[1], Yuxuan Xiao[2], Luke Wernert[3], Jeffrey A. Brock[2], Eric E. Fullerton[2], Hua Chen[3,4], Hailong Wang[2,*], and Chunhui Rita Du[1,2,*]

[1]Department of Physics, University of California, San Diego, La Jolla, California 92093
[2]Center for Memory and Recording Research, University of California, San Diego, La Jolla, California 92093-0401
[3]Department of Physics, Colorado State University, Fort Collins, Colorado 80523
[4]School of Advanced Materials Discovery, Colorado State University, Fort Collins, Colorado 80523

†These authors contributed equally to this work.

*Correspondence to: c1du@physics.ucsd.edu; h3wang@ucsd.edu

**Abstract**: Novel noncollinear antiferromagnets with spontaneous time-reversal symmetry breaking, nontrivial band topology, and unconventional transport properties have received immense research interest over the past decade due to their rich physics and enormous promise in technological applications. One of the central focuses in this emerging field is exploring the relationship between the microscopic magnetic structure and exotic material properties. Here, the nanoscale imaging of both spin-orbit-torque-induced deterministic magnetic switching and chiral spin rotation in noncollinear antiferromagnet Mn$_3$Sn films using nitrogen-vacancy (NV) centers is reported. Direct evidence of the off-resonance dipole-dipole coupling between the spin dynamics in Mn$_3$Sn and proximate NV centers is also demonstrated with NV relaxometry measurements. These results demonstrate the unique capabilities of NV centers in accessing the local information of the magnetic order and dynamics in these emergent quantum materials and suggest new opportunities for investigating the interplay between topology and magnetism in a broad range of topological magnets.

1. Introduction

Exploring and understanding new materials with advanced properties tied to their nontrivial electronic structures has been a central focus of modern condensed matter physics over the past few decades.[1–3] Noncollinear antiferromagnets based on the $Mn_3X$ composition (X = Sn, Ge, Ga, Ir, Pt, Rh) are particularly relevant in this context and have recently been the subject of immense research.[4–13] Due to the spontaneous time-reversal symmetry breaking, noncollinear magnetic order, and topologically nontrivial band structure near the Fermi surface, a range of counterintuitive and exotic phenomena have been observed in this emergent class of quantum materials. Examples include large anomalous Hall and anomalous Nernst effects,[8–10,12,13] chiral anomalies,[4] magnetic spin Hall effects,[14,15] and the promotion of long-range supercurrents,[16] offering an attractive platform to explore the interplay between topology, electron correlations, and magnetism. On another front, the noncollinear antiferromagnetic order hosted in $Mn_3X$-type compounds promises to deliver new data recording and storage functionalities, including terahertz-scale processing speeds, significantly improved stability, reproducibility, and ultra-high densities that outperform conventional ferromagnet-based spintronic technologies.[17–19]

Despite the enormous technological promise and rich fundamental physics of $Mn_3X$ compounds, the establishment of direct correlations between the micro- and mesoscopic magnetic structure and the intriguing properties of these materials remains elusive. These challenges are particularly apparent for nanometer-thick $Mn_3X$ films,[4–7,20–24] in which the vanishingly small magnetic flux arising from the nearly-compensated magnetization is difficult to access using existing bulk magnetometry methods. In most of the previous studies, the magnetic properties of $Mn_3X$ films were typically inferred indirectly from magneto-transport measurements[4,5,10] or with other global measurement techniques such as neutron scattering[25] and vibrating-sample magnetometry,[6,8,21] rendering limited information on their microscopic spin configurations. In this context, a reliable experimental approach capable of detecting the local magnetic behavior of $Mn_3X$ films at the nanoscale is highly desirable for a comprehensive understanding of this new family of topological magnets.

To address these challenges, we employ nitrogen-vacancy (NV) centers, optically active atomic defects in diamond that act as single-spin sensors,[26–31] to perform nanoscale quantum sensing and imaging of $Mn_3Sn$ films. Exploiting the NV wide-field magnetometry technique,[32–35] we directly accessed the local magnetic textures of $Mn_3Sn$ films at a sub-micrometer length scale and imaged the spin-orbit-torque (SOT)-driven[36] deterministic magnetic switching[4] and chiral spin rotation.[5,19] Employing NV relaxometry,[37–41] we observed the dipole-dipole coupling between the spin dynamics of $Mn_3Sn$ and proximate NV centers, providing an attractive solid-state platform to develop NV-based hybrid architectures for quantum technological applications. Our results highlight the significant potential of NV centers for probing the nanoscale magnetic properties of emergent magnetic topological materials with low moments,[42–47] and suggest new opportunities for investigating the interplay between inhomogeneous magnetic order and topological band structure in a broad range of quantum materials.

2. Results and Discussion
2.1. Antiferromagnetic topological semimetal $Mn_3Sn$

We first briefly review the pertinent material properties of $Mn_3Sn$. $Mn_3Sn$ is a hexagonal antiferromagnetic metal that hosts a noncollinear inverse-triangular spin configuration on the kagome lattice of the (0001) planes,[4,8] as illustrated in Figure 1a. The spin structure can be viewed as the ferroic ordering of a cluster magnetic octupole consisting of six non-collinearly ordered Mn atoms that reside in two stacked kagome layers. The characteristic inverse triangular spin configuration has an orthorhombic symmetry, and only one of the three moments in each Mn triangle is parallel to its local easy axis. Thus, canting of the other two Mn spins towards their local easy axes gives a vanishingly small remnant magnetization (~0.002 $\mu_B$ per Mn atom) below the Néel temperature of ~420 K.[8] The spontaneous time-reversal symmetry breaking induced by the noncollinear antiferromagnetic order enables the formation of Weyl nodes[2,48] near the Fermi surface and a non-vanishing Berry curvature in momentum space, leading to an exceptionally large topological magneto-transport (thermal) response.[8,9,11,15] In the present study, we employed 50-nm-thick polycrystalline $Mn_3Sn$ films prepared by ultra-high-vacuum magnetron sputtering (Section 1, Supporting Information). In agreement with previous studies,[4,8] magneto-transport measurements indicate that the anomalous Hall effect in our $Mn_3Sn$ samples is approximately half the magnitude observed in bulk crystals.[8] To achieve SOT-driven deterministic magnetic switching and chiral spin rotation, we deposited heavy metal layers (Pt or W) on top of the $Mn_3Sn$ films *in-situ* and patterned the samples into standard Hall cross devices with a width of 10 μm for electrical transport and NV measurements.

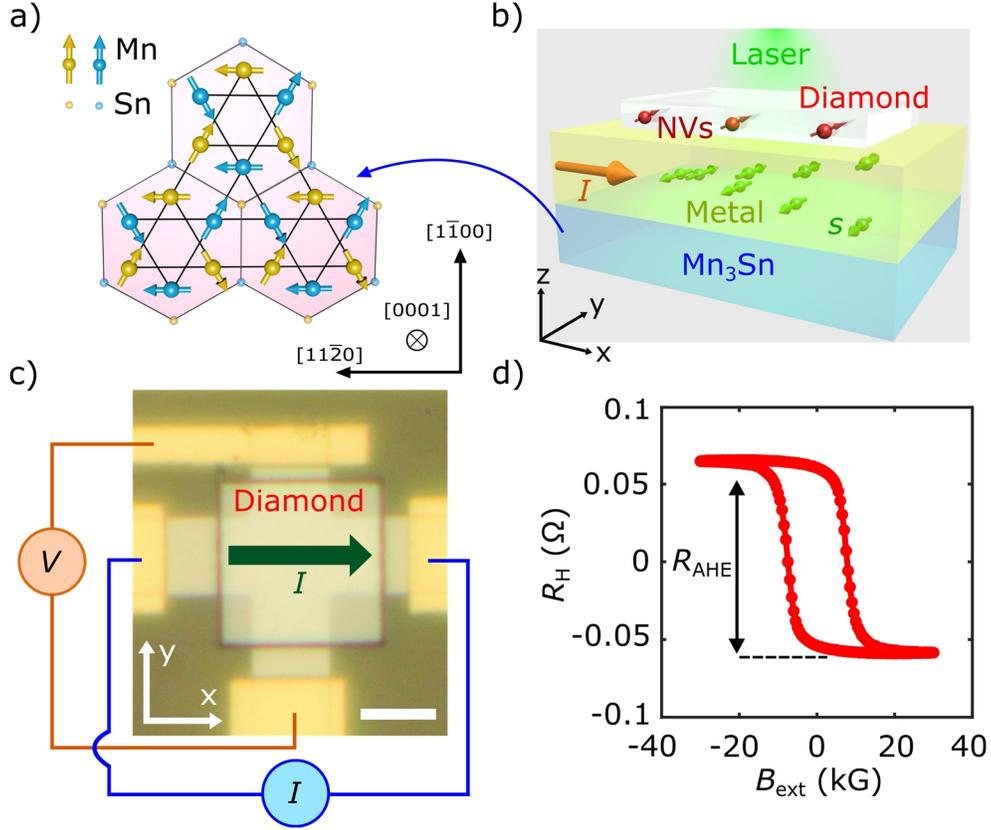

**Figure 1**. Antiferromagnetic Weyl semimetal $Mn_3Sn$ and bilayer device layout. a) Schematic of the kagome lattice of $Mn_3Sn$ hosting inverse triangular spin configurations, here the yellow and blue arrows (balls) represent the Mn atoms and Sn atoms located in different kagome layers. b) Schematic illustration of the nitrogen-vacancy (NV) wide-field magnetometry measurement platform, where a diamond microchip containing NV ensembles is positioned on top of a $Mn_3Sn$/metal sample. A charge current $I$ flowing through the metal layer generates a spin current (green arrows) via the spin Hall effect. The produced spin currents flow across the interface and exert spin-orbit-torques on the local magnetization of $Mn_3Sn$. c) Optical image of a prepared $Mn_3Sn$/Pt Hall device with an illustration of the magneto-transport measurement geometry. Electrical write and read currents are applied along the $x$-axis, and the Hall voltage is measured along the $y$-axis. The scale bar is 10 μm. d) Hall resistance $R_H$ of the $Mn_3Sn$/Pt device measured as a function of perpendicular magnetic field $B_{ext}$.

Nanoscale magnetic sensing and imaging was performed using NV centers as illustrated in Figure 1b. A diamond microchip[49] with lateral dimensions of 20 μm × 20 μm is positioned on top of a $Mn_3Sn$/metal Hall device. NV centers are shallowly implanted on the bottom surface of the diamond membrane with a density of ~1500/μm$^2$, providing a suitable platform for wide-field magnetometry measurements based on the ensemble of NV spins (Section 1, Supporting Information). Figure 1c shows an optical image of a prepared device, where the Au pads are arranged for standard four-probe Hall voltage measurements, and a diamond microchip placed on top of the patterned Hall cross is used to detect SOT-driven magnetic dynamics in $Mn_3Sn$. Figure 1d shows the anomalous Hall resistance of the $Mn_3Sn$/Pt Hall device measured as a function of the applied perpendicular magnetic field $B_{ext}$. The characteristic "negative" anomalous Hall resistance of $Mn_3Sn$ changes sign with the reversal of the external magnetic field, in agreement with previous studies.[4,8]

We first present data to demonstrate our ability to achieve SOT-driven deterministic magnetic switching of $Mn_3Sn$. In these measurements, millisecond write current pulses $I_{write}$ with a systematically varied magnitude are applied along the *x*-axis of the patterned Hall device. As illustrated in Figure 2a, $I_{write}$ generates a spin current that flows along the *z*-axis with spin polarization *s* along the *y*-axis via the spin Hall effect.[36,50] The spin current flows across the $Mn_3Sn$-metal interface and exerts field-like and damping-like SOTs on the chiral spin structure of the $Mn_3Sn$ layer. When an external static magnetic field is applied along the electric current direction, sufficiently large spin currents will drive deterministic switching of the cluster magnetic octupoles when the kagome planes are parallel to the spin polarization *s*,[4] as illustrated in Figure 2b. Figure 2c shows the SOT measurement results of a patterned $Mn_3Sn$/Pt Hall device. The measured Hall voltage signals show positive or negative jumps at the critical write currents depending on the sign of the external magnetic field.[4] As shown in Figure 2d, when the Pt is replaced by W, a material with a negative spin Hall angle,[51,52] the measured Hall signals show opposite switching polarity (Section 2, Supporting Information). By comparing the variation of the Hall signals in the current-induced switching experiments and the field-induced switching experiments (Figure 1d), we determine the proportion of flipped magnetic domains during switching to be ~25% and ~30% in the Pt- and W-based Hall devices, respectively.

## 2.2. Quantum imaging of SOT-driven deterministic magnetic switching in $Mn_3Sn$

We next utilized NV magnetometry to reveal the deterministic magnetic switching of $Mn_3Sn$ on the microscopic scale. Figure 3a shows the measured Hall resistance $R_H$ of a $Mn_3Sn$/Pt Hall device as a function of the write current $I_{write}$. At a series of points ("A" to "I") marked on the hysteresis loop, we performed NV wide-field magnetometry measurements after each of the Hall voltage measurements to visualize the SOT-induced variation of the magnetic stray fields. To give an intuitive picture, the SOT systematically changes the perpendicular magnetization of $Mn_3Sn$ in a current-induced magnetic hysteresis loop,[4] leading to a variation of the stray fields generated by the sample. Thus, by imaging the evolution of the stray field maps as a function of the write current, we can infer the variation of the local magnetic textures and interpret the underlying switching mechanism of Mn$_3$Sn.

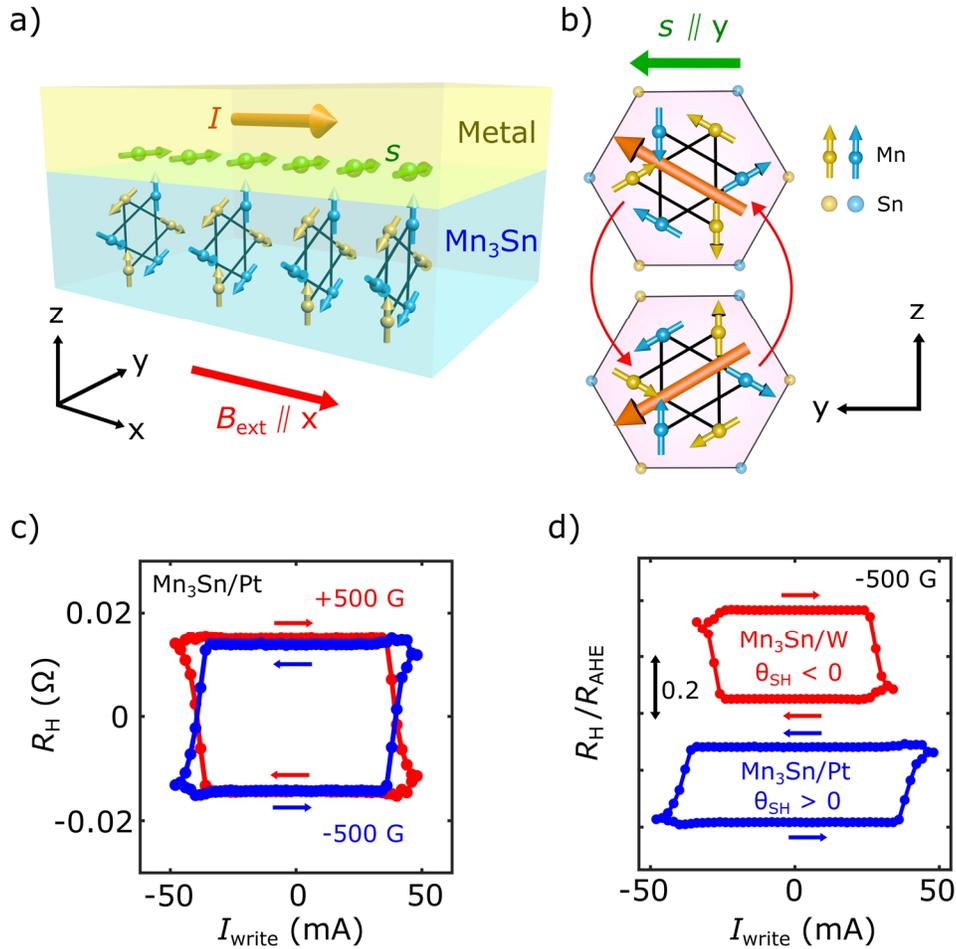

**Figure 2.** Electrical transport characterization of spin-orbit-torque (SOT)-driven deterministic magnetic switching in Mn$_3$Sn/metal Hall devices. a) Schematic illustration of SOT-driven deterministic magnetic switching of Mn$_3$Sn in the presence of an external magnetic field along the current direction (x-axis). The orange arrow represents the applied electric current and the green arrows represent the polarization of the generated spin currents. The characteristic kagome lattice consists of non-collinearly ordered Mn moments (blue and yellow balls and arrows represent the Mn atoms and their magnetic moments, respectively). b) An external magnetic field $B_{ext}$ along the electric current direction induces deterministic switching of the magnetic octupoles of Mn$_3$Sn with the kagome planes perpendicular to $B_{ext}$. c) Measured Hall resistance $R_H$ of a patterned Mn$_3$Sn/Pt Hall device as a function of the write current $I_{write}$ with $B_{ext}$ of 500 G (red curve) and −500 G (blue curve) applied along the current direction. d) Normalized Hall resistance $R_H/R_{AHE}$ for Mn$_3$Sn/Pt (blue curve) and Mn$_3$Sn/W (red curve) Hall devices, showing opposite magnetic switching polarities.

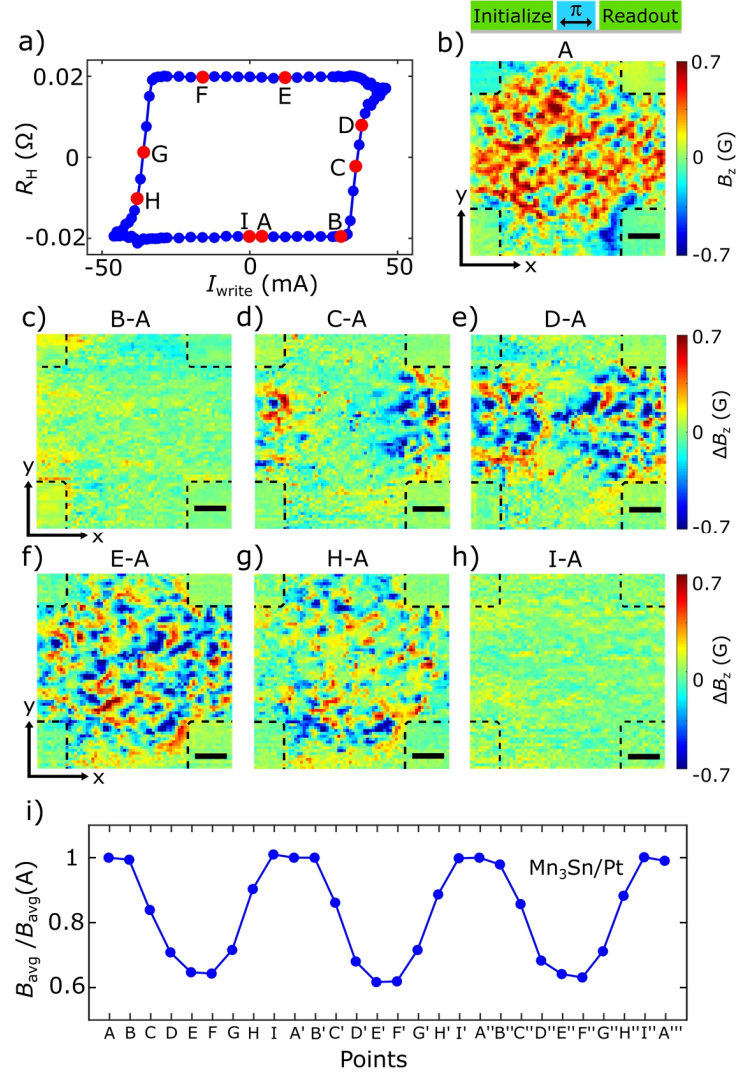

**Figure 3.** Quantum imaging of spin-orbit-torque (SOT)-driven deterministic switching of a $Mn_3Sn/Pt$ device. a) Measured Hall resistance $R_H$ as a function of write current $I_{write}$. NV wide-field imaging measurements are performed at individual points from "A" to "I" marked on the current-induced magnetic hysteresis loop. b) Two-dimensional (2D) image of the out-of-plane component of the magnetic stray field $B_z$ generated by the $Mn_3Sn$ sample in the initial state "A". The optical and microwave sequence used in the NV-based optically detected magnetic resonance measurements is depicted on top. c-h) 2D imaging of SOT-induced variation of the stray field $\Delta B_z$ at different states marked on the hysteresis loop. $B_z$ measured in the initial state "A" has been subtracted for visual clarity. For (b-h), black dashed lines mark the boundary of the patterned $Mn_3Sn/Pt$ Hall cross with a width of 10 μm, and the scale bar is 2.5 μm. i) SOT-induced variation of the spatially averaged stray field $B_{avg}$ normalized to the value $B_{avg}$ (A) measured in the initial state "A". Measurements were cycled through multiple magnetic hysteresis loops and the periodic variation in the behavior confirms the deterministic nature of the switching in $Mn_3Sn$. X, X', X", and X'" (X = A, B, C, D, E, F, G, H, and I) represent the same measurement point marked on the hysteresis loop at successive cycle numbers.

Before discussing the details of the measurement results, we first briefly discuss the quantum mechanical character of NV centers and the underlying principle of our measurements.

An NV center consists of a nitrogen atom adjacent to a carbon atom vacancy in one of the nearest neighboring sites of a diamond crystal lattice. The negatively charged NV state has an $S = 1$ electron spin and serves as a three-level quantum system.[26] NV wide-field magnetometry measurements exploit the Zeeman splitting effect of the ensemble of NV spins to measure the local static magnetic field along the NV-orientation.[26] The top panel of Figure 3b shows the measurement protocol. We utilized 1-μs-long green laser pulses for NV initialization and readout and ~100-ns-long microwave $\pi$ pulses[53] to induce NV spin transitions. The entire sequence is repeated ~15,000 times during one camera exposure period and the frequency $f$ of the microwave $\pi$ pulses is swept for different camera exposure periods.[33] When $f$ matches the NV electron spin resonance (ESR) conditions, NV spins are excited to the $m_s = \pm 1$ states, emitting reduced red photoluminescence (PL), which can be optically addressed. By measuring the splitting of the NV ESR energies via optically detected magnetic resonance measurements, the magnitude of the local magnetic field experienced by individual NV spins $B_{NV}$ can be quantitatively measured as: $B_{NV} = \pi(f_+ - f_-)/\gamma$,[26] where $\gamma$ is the gyromagnetic ratio and $f_\pm$ correspond to the NV ESR frequencies of the $m_s = 0 \leftrightarrow \pm 1$ transitions. Subtracting the contribution from the external magnetic field $B_{ext}$, the stray fields $B_{NV}^M$ generated by the Mn3Sn sample can be obtained. Note that the spatial components of the stray field $B_{NV}^M$ are linearly dependent in Fourier space,[54] allowing for extraction of the out-of-plane component $B_z$ (Section 3, Supporting Information). The bottom panel of Figure 3b presents a two-dimensional stray field $B_z$ map of the Mn3Sn/Pt Hall device measured in the initial magnetic state "A" of the hysteresis loop. Notably, the obtained $B_z$ map shows characteristic multidomain features on the length scale of hundreds of nanometers (ultimately limited by the spatial resolution of the wide-field imaging measurement). Because $B_z$ is dominated by stray fields generated by the perpendicular magnetization,[54] we expect the spatial distribution of magnetic domains in the prepared polycrystalline Mn3Sn film to largely follow a similar pattern, suggesting multiple orientations of the magnetic octupoles throughout the sample.

Building on the demonstrated NV wide-field magnetometry platform, we now visualize the SOT-driven deterministic magnetic switching of Mn3Sn at different write currents. Figures 3c-3h show a series of representative magnetic stray field $B_z$ maps taken at the corresponding points ("B" to "I") on the current-induced magnetic hysteresis loop. For visual clarity, $B_z$ of the initial magnetic state at point "A" has been subtracted in the presented images to highlight the SOT-induced relative variations. The few red and blue regions in Fig. 3c are due to noise/uncertainty in the measurements. When $I_{write}$ is below the critical value, the measured $B_z$ map remains nearly the same, as shown in Figure 3c, indicating that the spin-current-induced magnetic switching is negligible. Above the critical write current, the SOT overcomes the local magnetic anisotropy and begins to drive magnetic switching of individual magnetic domains,[4] leading to the variation of the measured stray fields shown in Figure 3d. The observed magnetic switching features originate in the current leads where the local current density is initially higher and then spread across the entire Hall cross region with increasing write current (Figures 3e and 3f). Inverting the write current into the negative regime leads to a reversal of the magnetic switching polarity (Figure 3g). Lastly, when sweeping the write current back to point "I", the Mn3Sn/Pt Hall device shows an identical stray field map and Hall signal compared to the initial magnetic state, as shown in Figure 3h, suggesting that the local domain structure is tied to the underlying microstructure of the sample.

The NV wide-field imaging results highlight the nonuniform, SOT-driven magnetic switching behavior in Mn3Sn. Due to the polycrystalline nature of the prepared Mn3Sn Hall device, we expect a random distribution of the polarization of the cluster magnetic octupoles in individual magnetic grains.[4] Microscopically, SOT-induced deterministic switching preferably occurs in the crystalline grains with kagome-plane orientation parallel to the polarization of the injected spin currents,[4,5] as illustrated in Figure 2a. Once the magnetic octupole is reversed in these grains, the domain wall cannot propagate to neighboring grains due to their differing magnetic orientations across the grain boundary. As a result, the magnetization reversal process in the polycrystalline Mn$_3$Sn sample manifests as the gradual rotation of individual crystalline grains as the write current increases, in contrast to conventional domain wall nucleation and propagation behaviors observed in ferromagnetic(ferrimagnetic) systems[55] (Section 4, Supporting Information). It is worth noting that the change in the stray field $B_z$ relative to that in point "A" shows a coexistence of regions with opposite signs during the magnetic switching process of Mn$_3$Sn. An immediate explanation is that it may be due to the uncompensated surface magnetization[56] ubiquitous to antiferromagnets[57,58]. Due to the random surface environments of the Mn$_3$Sn grains, the surface magnetization orientation may be opposite to that of the weak bulk magnetization, which may lead to a differing local sign of $B_z$ even if the bulk magnetic order of the grains is set deterministically by the SOT.

Because the measured stray field pattern is essentially a projection of the distribution of magnetic domain orientations, spatial averaging of $B_z$ over the entire Hall cross region serves as a qualitative indicator of the macroscopic magnetic state of the Mn$_3$Sn device. To further corroborate the observed SOT-driven magnetic switching, Figure 3i plots the variation of the spatially averaged perpendicular magnetic stray field $B_{avg}$ (normalized to the value $B_{avg}$ (A) measured in the initial state "A") at individual measurement points ("A" to "I"). Notably, $B_{avg}$ systematically varies as the write current is swept and returns to the original value after one complete switching loop. Repeating the SOT and NV wide-field magnetometry measurements over multiple switching cycles, $B_{avg}$ shows a robust periodic variation, demonstrating that the observed deterministic magnetic switching in Mn$_3$Sn is highly reproducible. We note that measured Mn$_3$Sn/W devices exhibit similar behavior but with opposite switching polarity (Section 5, Supporting Information), suggesting that artefacts such as Oersted fields and thermal/electrochemical effects do not play a significant role in our measurements.

### 2.3. Quantum imaging of SOT-driven chiral spin rotation of Mn$_3$Sn

In addition to the deterministic magnetic switching discussed above, SOT can also drive continuous rotation of the chiral spin structure of Mn$_3$Sn, as has been recently reported.[5] Figure 4a illustrates the relevant measurement geometry, where spin currents with polarization along the *y*-axis are generated by an applied charge current through the spin Hall effect. Above a critical current value, spin currents can overcome the local magnetic anisotropy and Gilbert damping to drive continuous rotation of the magnetic octupoles of Mn$_3$Sn when the kagome planes lie perpendicular to the spin polarization. The dynamics of the individual sublattice magnetic moments $m_i$ can be described by the coupled Landau-Lifshitz-Gilbert (LLG) equation as follows:[4,5]

$$\frac{\partial \boldsymbol{m}_i}{\partial t} = -\gamma \boldsymbol{m}_i \times \mathbf{B}_{\mathrm{eff},i} + \alpha \boldsymbol{m}_i \times \frac{\partial \boldsymbol{m}_i}{\partial t} - \frac{\gamma \hbar \theta_{\mathrm{SH}} J}{2 e M_s d} \boldsymbol{m}_i \times (\boldsymbol{m}_i \times \mathbf{s}) \qquad (1)$$

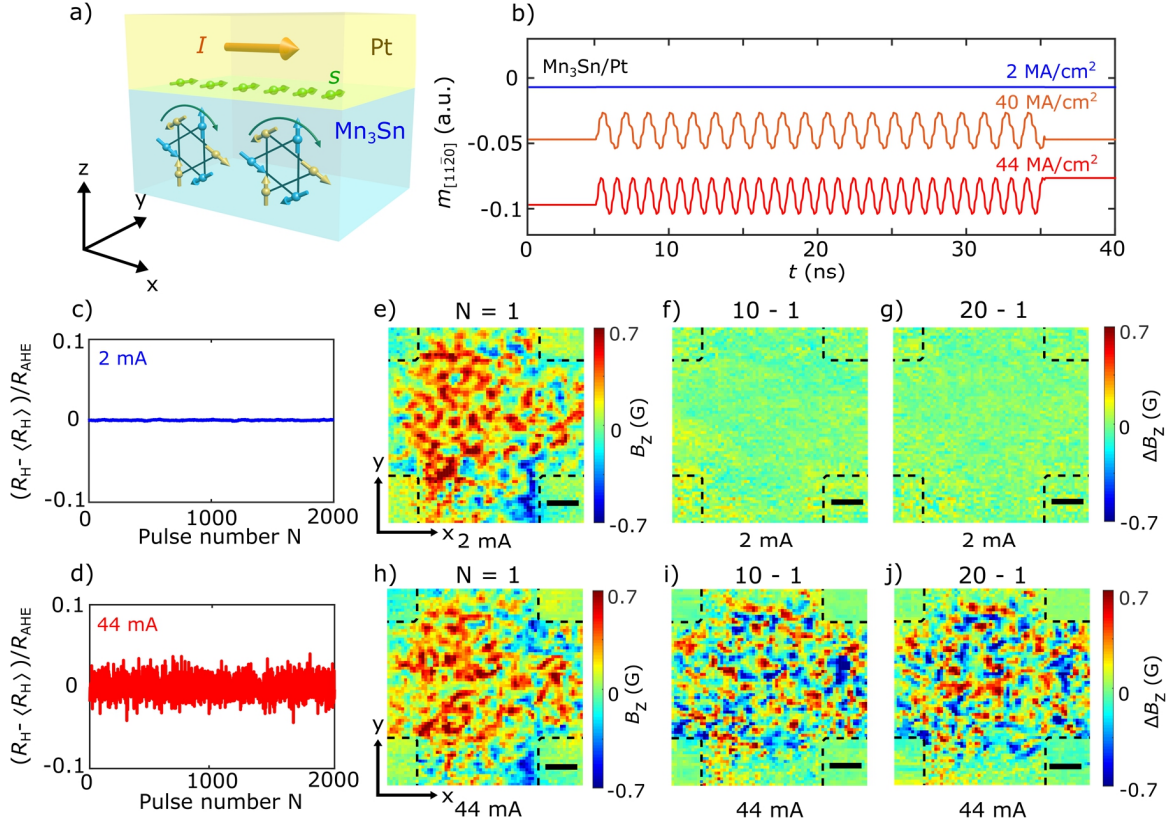

**Figure 4.** Quantum imaging of spin-orbit-torque (SOT)-driven chiral spin rotation in a Mn$_3$Sn/Pt device. a) Schematic illustration of the measurement geometry for SOT-driven chiral spin rotation in Mn$_3$Sn. Electric charge current flowing in the metal layer generates spin currents with polarization *s* along the *y*-axis. Above the critical current density, the generated spin current drives a continuous rotation of the magnetic octupoles of Mn$_3$Sn with the kagome planes perpendicular to *s*. The orange arrow represents the applied electric current, and the green arrows represent the polarization of the spin currents generated. b) Calculated time dependence of the weak ferromagnetic moment parallel to the [11$\bar{2}$0] axis ($m_{[11\bar{2}0]}$) for Mn$_3$Sn/Pt. The electric current is applied from $t = 5$ ns to $t = 35$ ns and the kagome plane is set to be perpendicular to the spin polarization axis *s* in the presented simulations. c,d) Measured normalized Hall resistance ($R_H - \langle R_H \rangle$)/$R_{AHE}$ as a function of pulse number (N) for write current amplitudes of 2 mA (c) and 44 mA (d). $\langle R_H \rangle$ is the average of $R_H$ and $R_{AHE}$ is the zero-field Hall resistance. e,h) 2D images of $B_z$ measured after applying the first current pulse (N = 1) with magnitudes of 2 mA (e) and 44 mA (h), respectively. f,g) SOT-induced variation of the stray field $\Delta B_z$ recorded after applying a 2 mA current pulse 10 (f) and 20 (g) times. i,j). SOT-induced variation of stray field $\Delta B_z$ recorded after applying a 44 mA current pulse 10 (i) and 20 (j) times. For (e-j), the black dashed lines mark the boundary of the patterned Mn$_3$Sn/Pt Hall cross, and the scale bar is 2.5 μm.

where $\hbar$ is the reduced Planck constant, $\theta_{SH}$ is the spin Hall angle of the heavy metal layer, $e$ is the electron charge, $M_s$ and $\boldsymbol{B}_{eff} = -(M_s)^{-1}\delta u/\delta \boldsymbol{m}_i$ are the magnetization and the effective magnetic field of each magnetic sublattice, $u$ is the magnetic energy density, $\alpha$ and $d$ are the Gilbert damping constant and the thickness of the Mn$_3$Sn film, and $J$ is the electric current density flowing in the heavy metal layer. Substituting the relevant material parameters and solving the above equation numerically, Figure 4b shows the time-dependent variation of the weak ferromagnetic moment (average magnetic moment of all sublattices) projected along the [11$\bar{2}$0] axis ($m_{[11\bar{2}0]}$) (Section 6, Supporting Information). In the presented simulations, the kagome plane is perpendicular to the spin polarization *s* and the electric current is on from $t = 5$ ns to $t = 35$ ns; we are mainly interested in SOT-driven magnetic dynamics during this period. When $J = 2$ MA/cm$^2$, $m_{[11\bar{2}0]}$ has a constant value, indicating a negligible effect of SOT on the magnetic state of Mn$_3$Sn. When $J = 40$ MA/cm$^2$ or 44 MA/cm$^2$, $m_{[11\bar{2}0]}$ exhibits periodic oscillations whilst the electric current is applied (*i.e.*, for 5 ns ≤ $t$ ≤ 35 ns). When the electric current is turned off, the chiral spin structure stops rotating and is expected to randomly settle into one of the degenerate easy magnetic configurations available in each crystalline grain. Cycling the above processes and monitoring the Hall response after individual SOT-driven chiral spin rotations, we would expect a characteristic fluctuation of the measured Hall signals around the half-level with the amount of fluctuation depending on the number of domains in the Hall cross region.[5]

We now present electrical transport measurement results supporting the physical picture discussed above. Figures 4c and 4d show the measured anomalous Hall resistance ($R_H - \langle R_H \rangle$) (normalized by the zero-field Hall resistance $R_{AHE}$) as a function of the write current pulse number N. The measurement is performed 2,000 times repeatedly for two amplitudes of $I_{write}$ below and above the threshold value. One can see that the larger write current ($I_{write} = 44$ mA) induces significantly enhanced fluctuations of the measured anomalous Hall signal, which is absent in the low-current case ($I_{write} = 2$ mA). Invoking a theoretical model presented in Ref. 5, the characteristic length scale of the magnetic domains in the Mn$_3$Sn sample is estimated to be ~300 nm, in qualitative agreement with our NV measurement results (Section 7, Supporting Information).

Next, we use NV wide-field magnetometry to establish a direct, real-space correlation between SOT and the chiral spin rotation of Mn$_3$Sn. Figures 4e and 4h show the stray field maps measured after applying the first write current pulse with magnitudes of 2 mA and 44 mA, respectively. Both maps exhibit characteristic multidomain features, consistent with the results shown in Figure 3b. We repeat the NV measurements after every electrical write current pulse to visualize the effects of SOT on the magnetic domain structure. Figures 4f, 4g, 4i, and 4j present the variation of the magnetic stray field maps recorded after applying the write current pulse 10 and 20 times with amplitudes of either 2 mA or 44 mA. The stray field measured after applying the first write current pulse has been subtracted for visual clarity. For $I_{write} = 2$ mA, the spatial profile of the stray fields largely remains unchanged when repeating the write current pulse (Figures 4f and 4g). In sharp contrast, the stray field images measured with $I_{write} = 44$ mA show substantial random variations, highlighting the characteristic SOT-driven continuous spin rotation in Mn$_3$Sn. While the local magnetic stray field pattern exhibits dynamic variations induced by individual write current pulses of 44 mA amplitude, the stray field spatially averaged over the entire Hall cross region remains constant during the whole measurement sequence (Section 8, Supporting Information). Such experimental results were expected because no in-plane bias magnetic field was applied in the chiral spin rotation measurements, and hence, SOT-driven deterministic magnetic switching should not occur.[4,5] Note that potential artefacts induced by Joule heating have been systematically excluded via control experiments (Section 9, Supporting Information).

## 2.4. Quantum sensing of SOT-driven continuous chiral spin rotation

In the previous section, we employed NV wide-field magnetometry to probe the SOT-driven magnetic switching in Mn$_3$Sn. Theoretically, a continuous rotation of the chiral spin order is expected under a constant electric current that will generate microwave magnetic fields whose frequency depends on the magnitude of the current.[5] Based on the theoretical model highlighted above, Figure 5a plots the calculated rotation frequency $f$ of the $Mn_3Sn$/Pt sample as a function of the electric current density $J$. Note that the threshold value of $J$ is calculated to be 36 MA/cm$^2$, in agreement with the value obtained from the NV wide-field magnetometry measurements (Section 10, Supporting Information). On the microscopic scale, individually rotating crystalline grains of Mn$_3$Sn serve as local resonant magnetic oscillators,[19] which can couple with proximate NV centers via the dipole-dipole interaction.[41,59] Next, we introduce NV relaxometry measurements[37–41] to reveal the coherent spin dynamics of Mn$_3$Sn, providing another piece of evidence for SOT-induced chiral spin rotation.

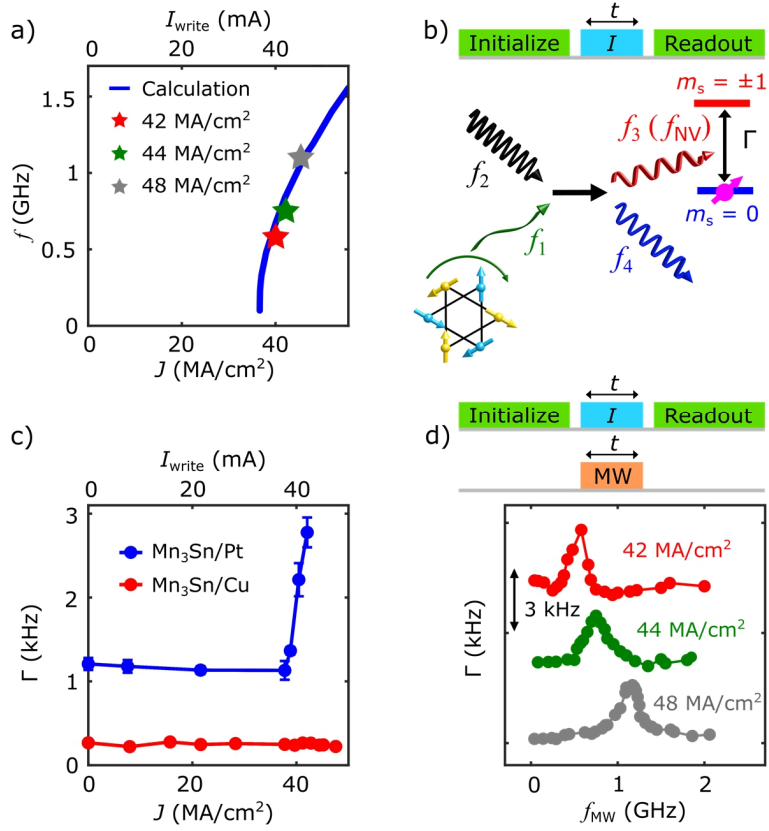

**Figure 5.** Quantum sensing of spin-orbit-torque (SOT)-driven chiral spin rotation of a $Mn_3Sn$/Pt device. a) Theoretically calculated spin rotation frequency $f$ of $Mn_3Sn$ as a function of the electric current density $J$ flowing in the Pt layer. The experimentally measured $f$ for $J = 42$ MA/cm$^2$ (the red marker), 44 MA/cm$^2$ (the green marker), and 48 MA/cm$^2$ (the grey marker) agree with the theoretical calculations. The experimental results are presented in Figure 5d. b) Top panel: Measurement protocol of NV relaxometry. Bottom panel: Schematic illustration of multimagnon-scattering mediated coupling between the chiral spin rotation of $Mn_3Sn$ and the NV centers. The SOT-induced continuous spin rotation of $Mn_3Sn$ generates magnons at frequency $f_1$. The produced magnons scatter with thermal magnons of $Mn_3Sn$ at frequency $f_2$ (black), generating two new magnons with frequencies $f_3$ (red) and $f_4$ (blue), as illustrated in the schematic. Circulating such four-magnon scattering processes leads to a new quasi-equilibrium state with an enhanced magnon density at the NV electron spin resonance frequency $f_{NV}$, resulting in accelerated NV spin relaxation. c) Measured NV relaxation rate $\Gamma$ as a function of the electric current density $J$ in the $Mn_3Sn$/Pt Hall device (blue) and a $Mn_3Sn$/Cu Hall device (red). The error bar of the measured $\Gamma$ in the control experiment is smaller than the marker size. d) Top panel: Optical and microwave sequence to measure the microwave-assisted chiral spin rotation of $Mn_3Sn$. Bottom panel: Measured NV relaxation rate $\Gamma$ as a function of the external applied microwave frequency $f_{MW}$ for $J = 42$ MA/cm$^2$ (red), 44 MA/cm$^2$ (green), and 48 MA/cm$^2$ (grey). The measured relaxation rate reaches a maximum value when $f_{MW}$ matches the chiral spin rotation frequency $f$ of $Mn_3Sn$. The obtained chiral spin rotation frequency $f$ at the corresponding electric current density agrees with the theoretical calculations shown in Figure 5a. The experimental error bar of the data presented in Figure 5d is smaller than the marker size.

We first discuss the procedure and principle of NV relaxometry measurements. Our measurements were performed on a Mn$_3$Sn/Pt Hall device with electric current densities $J$ ranging from 0 to 48 MA/cm$^2$. We found that the devices are prone to electrical breakdown for $J$ values greater than 53 MA/cm$^2$. The top panel of Figure 5b shows the protocol of the NV relaxometry measurement. A 3-µs-long green laser pulse is first applied to initialize the NV spins to the m$_s$ = 0 state. Next, an electric current pulse with a duration varying from 0 to 4 milliseconds is used to drive the local chiral spin rotation of Mn$_3$Sn. When the resonant frequency of the excited spin rotation is lower than the NV ESR frequency (in the electric current density range accessible to us), the low-frequency magnons generated by the chiral rotation will interact with the thermal magnon bath of Mn$_3$Sn through multimagnon scattering processes,[39,41,60] establishing a new quasi-equilibrium state with increased magnon density at the NV ESR frequency $f_{NV}$, as illustrated in the bottom panel of Figure 5b. The enhanced magnon population at frequency $f_{NV}$ will generate fluctuating stray fields, which induce NV spin transitions from the m$_s$ = 0 to the m$_s$ = $\pm$1 states that can be optically addressed through their spin-dependent photoluminescence. After a delay time $t$, we measure the NV occupation probabilities with a green-laser readout pulse. By measuring the integrated intensity of the NV photoluminescence as a function of the delay time $t$ and fitting the data with an exponential decay function, the NV relaxation rate Γ can be quantitatively obtained (Section 11, Supporting Information). Figure 5c shows the obtained NV relaxation rate Γ measured as a function of the electric current density $J$. When $J < 38$ MA/cm$^2$, one can see that Γ is roughly a constant value attributable to the NV's intrinsic spin relaxation rate. In agreement with the physical picture of multimagnon scattering-induced NV relaxation presented above, Γ shows significant enhancement when $J \geq 38$ MA/cm$^2$. Qualitatively, a larger electric current drives faster chiral spin rotation in the Mn$_3$Sn and "pumps" more thermal magnons at the NV ESR frequency. Therefore, we observe a monotonic increase of the measured NV relaxation rate when $J$ is above the critical value. The critical electric current density obtained experimentally agrees with the theoretical prediction (Figure 5a) and NV wide-field magnetometry results (Section 10, Supporting Information), confirming that the observed NV relaxation indeed results from the coherent spin dynamics of Mn$_3$Sn. We highlight that the current dependent enhancement of the NV relaxation rate is absent when replacing Pt with Cu (*i.e.*, a material with a vanishingly small spin Hall angle), demonstrating the central role of spin currents in driving magnetic dynamics in Mn$_3$Sn. Note that the potential effects of Joule heating have been carefully examined and excluded via control experiments (Section 9, Supporting Information).

To access the spin dynamics of Mn$_3$Sn in a more direct fashion, we utilize external microwave magnetic fields to enhance the coherence of local spin rotation in individual magnetic grains of Mn$_3$Sn. Specifically, we drive the chiral spin rotation of Mn$_3$Sn at a fixed electric current density and perform the NV relaxometry measurements under the application of an external microwave magnetic field, as illustrated by the measurement protocol shown in the top panel of Figure 5d. When the frequency $f_{MW}$ of the external microwave field matches the rotation frequency $f$ of Mn$_3$Sn, it effectively synchronizes the magnetic oscillations of individual Mn$_3$Sn grains due to the linear coupling between the oscillation modes and the spatially coherent microwave field[59,61,62] (Section 12, Supporting Information). The resonance also leads to enhanced oscillation amplitudes and increased fluctuations of the magnetic fields at the NV sites. The bottom panel of Figure 5d shows the measured NV relaxation rate Γ as a function of $f_{MW}$. We note that when $f_{MW}$ matches the SOT-driven chiral spin rotation frequency of $Mn_3Sn$, we observe a peak in the measured NV relaxation rate, demonstrating microwave-assisted coherent spin rotation of $Mn_3Sn$. The measured rotation frequencies agree with theoretical predictions based on the coupled LLG equation shown in Figure 5a.

## 3. Conclusion

In summary, we have applied NV-based quantum sensing and imaging techniques to investigate the microscopic magnetization dynamics in polycrystalline $Mn_3Sn$ films. By employing wide-field imaging techniques, we observed nonuniform, SOT-driven, deterministic magnetic switching behavior in $Mn_3Sn$ with a characteristic length scale of hundreds of nanometers. Using NV-based static field imaging and relaxometry methods, we provide mutually supporting evidence to experimentally corroborate the continuous chiral spin rotation of $Mn_3Sn$ in real space. Our results illustrate the unique capabilities of NV centers in diagnosing the local magnetic properties of emergent quantum materials with vanishingly small magnetic dipole moments. The demonstrated coupling between the low-frequency spin wave modes hosted by $Mn_3Sn$ and NV centers highlights the possibility of establishing and engineering mutual communication between topological magnets and quantum spin processors, opening new opportunities for developing hybrid functional architectures for next-generation, solid-state-based quantum information sciences and technologies.[63,64]

**Acknowledgements**
The authors would like to thank Nathan McLaughlin and Eric Lee-Wong for assistance in sample preparation. The authors thank Claudia Felser, Binghai Yan, Anastasios Markou, and Shunsuke Fukami for insightful discussions. G. Q. Y., S. L., M. H., and C. R. D. acknowledged support from U. S. National Science Foundation (NSF) under award DMR-2046227 and ECCS-2029558. H. L., H. W., and C. R. D. were supported by the Air Force Office of Scientific Research under award FA9550-20-1-0319 and its Young Investigator Program under award FA9550-21-1-0125. Y. X., J. A. B., and E. E. F were supported as part of Quantum Materials for Energy Efficient Neuromorphic Computing, an Energy Frontier Research Center funded by the U.S. DOE, Office of Science (Award No. DE-SC0019273). L. W. and H. C. were supported by the National Science Foundation CAREER Grant DMR-1945023.